\documentclass[doublecol]{epl2} 
\usepackage{amsmath}
\usepackage{graphicx}
\usepackage{bbold}
\usepackage{braket}
\usepackage{physics,color}
\newcommand{\md}[1]{\textcolor{black}{#1}}
\newcommand{\mga}[1]{\textcolor{black}{#1}}
\title{Are intrinsic decoherence models physical theories?}
\author{Maria Danelli\inst{1} and Matteo G. A. Paris\inst{1}}
\institute{\inst{1}Dipartimento di Fisica {\em 
Aldo Pontremoli}, Universit\`a di Milano, I-20132 Milano, Italy}
\abstract{Intrinsic decoherence models (IDMs) have been proposed in order to solve the measurement problem in quantum mechanics. In this work, we assess the status of two of these models as physical theories by establishing the ultimate bounds on the estimability of their parameters. Our results show that dephasing and dissipative IDMs are amenable to falsification and should be considered physical theories worthy of experimental study.}
\begin{document}
\maketitle
\section{Introduction}
\md{Quantum mechanics is a well-established and 
self-consistent theory that has repeatedly aligned with experimental results. However, certain conceptual challenges remain under discussion, particularly regarding the description of macroscopic objects and their interactions with microscopic systems. These challenges often relate to the presence of linear superpositions involving macroscopically distinguishable states in macroscopic systems \cite{Lombardi17,benatti03}. }
Decoherence of macroscopic systems is often attributed to their interaction 
with the external environment. \md{When a system interacts with another that is subsequently traced out, it loses coherence and transitions into a statistical mixture, effectively resembling a classical system. As macroscopic systems are inherently coupled, this framework aptly explains the decoherence observed in the macroscopic world.} Yet, one may not be entirely satisfied with this explanation because it applies only to subsystems, {\em i.e.}, an external environment is always required, thus it still leaves open the question of how a system that is not part of a bigger system, such as the entire universe, would behave.  This lack of complete satisfaction led to searching for other possible explanations, or at the very least, it remains uncertain whether alternative explanations can be unequivocally 
ruled out.

A possible explanatory path is represented by intrinsic decoherence models  (IDMs) which try to soften the strong clash between the Schr\"odinger's equation and the reduction postulate \cite{leg02,stamp12}. In fact, in quantum 
mechanics when a measure is done the wave function of the system collapses, and how this collapse happens needs to be described within quantum 
theory. But how could this stochastic, non-unitary, non-linear phenomenon be described by the deterministic, unitary, linear temporal evolution governed by Schr\"odinger's equation? \md{If, modifying 
Schr\"odinger's equation, decoherence emerges from temporal evolution, the reduction postulate is no longer required, thereby eliminating the need to resolve the contradiction.}

This idea was discussed in \cite{4,kanai48} where energy dissipation was linked 
to an explicit time dependent Hamiltonian, corresponding to a Lagrangian describing the exact classical damped evolution equation. In \cite{ALBRECHT1975127}, a class of friction potentials was suggested, and approaches based on  the Schrodinger-Langevin \cite{dek81} or
Hamilton-Jacobi equation has been later presented \cite{STOCKER1979436}. In \cite{9} a different approach was suggested, introducing  a non-linear 
temporal evolution for state vector, thus without requiring a classical potential to quantize, but developing the model directly on the Hilbert 
space. We will focus on this model as it involves a purely dissipative mechanism to describe decoherence. The non-linear approach has been 
further explored by a field theory aimed at incorporating causality \cite{13} and adding nonlinear terms \cite{disentanglement} to describe 
disentanglement in macroscopic systems. 

The implementation of intrinsic decoherence within a non-dissipative and linear framework has been explored through the mechanism of spontaneous 
localization  \cite{1}. More recently, gravity has also been proposed as a means to achieve the same effect \cite{Gravity}. A paradigmatic 
model of dephasing-induced decoherence has been introduced in \cite{2} by modifying the von Neumann equation for the density operator. A modification 
to this model in order to make it Lorentz invariant has been developed \cite{Milburn2003}. The model of \cite{2} has been more recently analyzed to discuss its implication in quantum information and communications protocols \cite{2024,applied,14}. Further discussions about the validity and the implications of intrinsic decoherence models may be found in \cite{hsiang2022no,wu2017intrinsic,hillier2015continuous,gooding2014self,anastopoulos2008intrinsic,diosi2005intrinsic,bonifacio2002puzzling,rajagopal1996implications,finkelstein1993comment,milburn1993reply,Alenezi2022}.

The aim of this work is to contribute to the discussion
using tools from quantum estimation theory. In particular, 
we focus on two paradigmatic IDM models involving pure dephasing and pure dissipation and evaluate the ultimate bounds to the estimability of their parameters. In this way, we investigate whether those models may be falsified or not, therefore assessing their status as physical theories. \md{The dephasing dynamics of MID have been associated with a potential stochastic unraveling driven by a quantum jump process \cite{Adler_2002,Brody_2023}. If these predictions differ significantly from experimental observations, this type of dynamics could potentially be ruled out.} Here, we aim  at investigating more fundamental limitations, by establishing the ultimate bounds to the quantum signal-to-noise ratio, which quantifies the inherent 
estimability of IDM parameters. An affirmative answer to our question would pave the way for conducting experiments on them in order to exclude, or even validate, one of the solutions to the measurement problem. Our results suggest that 
there exist optimal conditions in which it is realistic to design estimation protocols, therefore making IDMs falsifiable. These optimal conditions 
involve setting the effective evolution time of the system under investigation of the order of magnitude of the inverse of the parameters. Additionally, 
it is required to select appropriate initial conditions,  e.g. choosing initial preparations with maximum coherence.

The paper is structured as follows. In the next section, we review the intrinsic decoherence models we are going to investigate,  {\em i.e.}, intrinsic 
dephasing and intrinsic dissipation. We then briefly 
review the tools of quantum estimation theory, whereas in the following two Sections \ref{s:deph} we analyze the two models in some detail, by applying them to a two-level system and to a harmonic oscillator and evaluating the ultimate bound on the estimability of the intrinsic decoherence parameters.  We also seek the optimal conditions that would give us the possibility to reach this ultimate bound on precision. The last Section closes the paper with some concluding remarks.
\section{Intrinsic decoherence models}\label{s:idm}
Intrinsic decoherence models reinterpret phenomena that are usually attributed to the interaction with the external environment attributing them to \textit{intrinsic} reasons, such as temporal evolution. The main phenomena experienced by coupled systems are 
dephasing,  {\em i.e.}, the loss of coherence, and dissipation,  {\em i.e.}, the loss of energy. 
In the following, we will focus on two different paradigmatic models of intrinsic 
decoherence, each one focusing on one of these phenomena. These models 
modify Schr\"odinger's equation to allow these effects to emerge naturally 
with temporal evolution. By doing so, they eliminate the need for the 
reduction postulate and solve the contradiction between deterministic 
evolution and stochastic measurement. The two approaches are complementary: 
the first one assumes a stochastic evolution and leads to dephasing, while in the other one evolution is deterministic but nonlinear and induces 
dissipation.   

\subsection{Intrinsic dephasing}
To describe intrinsic dephasing we employ the linear, 
stochastic model proposed by Milburn \cite{2} following previous approaches \cite{bonifacio1983coarse,3}. The basic assumption is that on 
sufficiently short time steps, the system 
does not evolve continuously under unitary evolution but rather in a stochastic 
sequence of identical unitary transformations \cite{bonifacio1983coarse,2},  {\em i.e.}, a minimum time step 
in the evolution of the universe is established. This yields surprising consequences, for instance, oscillatory systems become frozen at very high energies as it is impossible to produce an oscillator with a period shorter than this
minimum time step. This minimum time increment serves as an expansion parameter, with its inverse representing the mean frequency of the unitary step.  If this frequency is large enough, the system's evolution manifests as approximately continuous on laboratory time scales resulting in the following first-order 
master equation 
\begin{equation} 
    \frac{d}{dt}\rho (t) = -i [H, \rho]- \mu [H,[ H, \rho]]
    \label{mil}
\end{equation}
where we have assumed $\hbar=1$, and $\mu$ denotes the intrinsic dephasing parameter (it has the dimension of a time).
\md{Notice that implementing this kind of equation for fundamental dynamics does not detectably conflict with experimental data for prototypical quantum systems \cite{Adler_2002}. }
\subsection{Intrinsic dissipation}
In this section, we explore deterministic, dissipative, non-linear models 
for quantum evolution, which are often used to describe transitions between 
atomic energy levels or more broadly, scenarios involving irreversible 
energy exchange between two systems. It is well-established that for these 
deterministic models to induce decoherence, non-linearity is required 
\cite{6}. However, the combination of non-linearity and determinism introduces a challenge, as it can lead to issues such as signaling \cite{7}. 
This apparent contradiction was only recently addressed in \cite{caban}, 
where it was demonstrated that by considering the master equation ($\gamma$ is an adimensional quantity)
\begin{equation}
\frac{d}{dt}\rho (t) = -i [H, \rho]+ \gamma\{H, \rho\} 
- 2 \gamma \rho \hbox{Tr}(H \rho)
\label{masterpolacca}
\end{equation}
the evolution is actually non-linear but maintains unaltered linear 
superposition. In other words, the equivalence of ensembles is 
preserved in time and signaling is avoided.  If we now consider 
Eq. (\ref{masterpolacca}) in the specific case of an initial  
pure state, we find out that it corresponds to the evolution 
equation put forward by Gisin \cite{9} to describe intrinsic 
dissipation,  {\em i.e.}, 
\begin{equation} 
\frac{d}{dt}\rho (t)  = -i [H, \rho]- \gamma[[H, \rho],\rho].
\label{mastergisin}
\end{equation}
Since the subset of pure states is invariant under (\ref{masterpolacca}), a
counterpart of this equation for the state vectors may be derived. 
However, it is not uniquely determined by (\ref{masterpolacca}). In fact, 
the whole family 
\begin{equation}
      \frac{d}{dt}\ket{\psi} = ( -i H - \gamma H+ \gamma \frac{\mel{\psi}{H}{\psi}}{\bra{\psi}\ket{\psi}}  {\mathbb{1}} + i k  {\mathbb{1})} \ket{\psi}
      \label{statevector}
\end{equation}
depending on $k$, a real function of $\ket{\psi}$,  leads to the evolution equation (\ref{masterpolacca}) for the density matrix.  In particular, for $k=0$, we obtain 
the equation for the evolution of pure states suggested in \cite{9}, which we are going to address in this paper as a paradigmatic model 
of intrinsic dissipation. The model is non-linear, deterministic, and dissipative, without being plagued by signaling.

The formal solution of Eq. (\ref{statevector}) for k=0 is given by
\begin{equation}
\ket{\psi(t)} = \frac{e^{-(i+ \gamma ) Ht} \ket{\psi_0}}
{\sqrt{ \mel{\psi_0}{e^{- 2\gamma H t}}{\psi_0}}} 
\label{formal}
\end{equation}
\section{Quantum Estimation Theory}
\label{s:qet}
In this section, we briefly review quantum estimation theory \cite{Paris09} and 
the tools to assess the status of the models discussed in the previous Section as 
part of physical theories.  Given a statistical variable $X$ with 
outcomes distributed according 
to $p(x|\lambda)$, we may estimate the value of $\lambda \in \mathbb{R}$ by performing 
repeated measurements of $X$ and processing data by a suitable estimator, 
 {\em i.e.} a function from the data space to the real axis. The precision of this 
estimation strategy is quantified by the variance of the estimator. 
For unbiased estimators the variance is bounded from below by 
the Cramer-Rao bound
$  V(\lambda) \geq [M F(\lambda)]^{-1}$, where $V$ denotes the variance of the estimator, $M$ is the number of measurements and $F$ is the so-called Fisher information (FI),  {\em i.e.} 
 \begin{equation}
     F(\lambda) = \int \frac{1}{p(x|\lambda)} \left( \frac{\partial  p(x|\lambda)}{\partial {\lambda}}\right)^2 dx\,.
\label{Fisher}
 \end{equation}
 In a quantum setting, the conditional distribution is given by the Born rule 
 $p(x|\lambda)=\hbox{Tr}[\rho_\lambda\, \Pi_x]$, where $\varrho_\lambda$ is the $\lambda$-dependent state of the system under investigation and $\{\Pi_x\}$ is the probability operator-valued measure (POVM) describing the measurement apparatus. 
 
Upon maximizing the FI over all the possible POVMs, it is possible to bound the Fisher Information (FI) by the quantum Fisher information (QFI)
\begin{equation}
        F(\lambda) \leq \hbox{Tr}(\rho_{\lambda} L_{\lambda}^2) \equiv Q(\lambda)  = \hbox{Tr}(\partial_{\lambda}\rho_{\lambda} L_{\lambda})
        \label{H}
\end{equation}
where $L_{\lambda}$ is the so-called Symmetric Logarithmic Derivative (SLD) defined as the selfadjoint operator satisfying the equation
$
  L_{\lambda} \rho_{\lambda} + \rho_{\lambda} L_{\lambda} = 2 \partial_\lambda \rho_{\lambda}{\partial{\lambda}}$. 
The Quantum Crao-Rao bound thus reads
$V(\lambda) \geq [M Q\lambda)]^{-1}$.
Notice that in this way the ultimate bound to precision depends only on the features 
of the \textit{quantum statistical model} $\rho_{\lambda}$.

We must now acknowledge that the smaller is the value of the parameter to be estimated, the higher should be the required precision. Therefore, the crucial quantity to consider is the scaling of the variance relative to the mean value of the parameter. It is useful then to define the \textit{signal-to-noise ratio} $U_{\lambda}$. 
\begin{equation}
    U_{\lambda} = \frac{\lambda^2}{V(\lambda)}
\end{equation}
The larger is $U_{\lambda}$, the better is the estimator. Using the Cramer-Rao bound, it is possible to derive the bound for $U_{\lambda}$, given by the \textit{quantum signal-to-noise ratio} (QSNR) $R_\lambda$
\begin{equation}
     U_{\lambda} \leq R_{\lambda} \equiv \lambda^2 Q(\lambda) 
\end{equation}
A large QSNR means that the parameter can in principle be estimated with high precision, whereas a small QSNR indicates that the value of the parameter is inherently uncertain, 
whatever strategy is employed for its estimation.  
 
We are now going to quote some explicit formulae for $Q(\lambda)$. Primarily, it is required to solve the equation for $L_{\lambda}$. Expressing $\rho_{\lambda}$ in his eigenbasis $\rho_{\lambda} = \sum_n \rho_n \ket{\psi_n} \bra{\psi_n}$, one obtains 
\begin{equation}
    Q(\lambda) =2 \sum_{m n} \frac{|\mel{\psi_m}{\partial_{\lambda} \rho_{\lambda}}{\psi_n}|^2}{\rho_m+ \rho_n} \,,
    \label{numerico}
\end{equation}
where the sum includes only terms with $\rho_m+ \rho_n \neq 0$. For pure states, the QFI can be expressed as 
\begin{equation}
 Q(\lambda) = 4\left(\bra{\partial_{\lambda}{\psi{\lambda}}} \ket{\partial_{\lambda}{\psi_{\lambda}}} -\left|\bra{\psi_{\lambda}} \ket{\partial_{\lambda}{\psi_{\lambda}}}\right|^2\right) 
 \label{QFIpuri}
\end{equation}

\section{Metrology of the MID parameter} \label{s:deph}
We are now going to apply the linear, stochastic, dephasing model of 
temporal evolution presented by Milburn (MID model) to a two-level 
system and a harmonic oscillator and observe the resulting dynamic. 
Afterward, we are going to study, by calculating the Quantum Fisher 
Information of these systems, if the influence of the decoherence 
parameter $\mu$ is detectable. 
\subsection{Two-level system}
\begin{figure}[ht!]
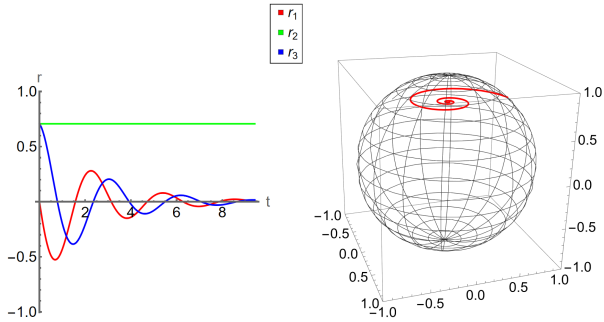

\centering
\includegraphics[width=0.44\columnwidth]{f1a_RayM.pdf}
\includegraphics[width=0.44\columnwidth]{f1b_sferamilb.pdf}
\caption{Panel (a): Temporal evolution of the components of 
the Bloch vector  of a two-level system with $H= \omega \sigma_3$, 
according to MID model, with initial condition $\theta= \frac{\pi}{4}$ 
and $\phi=\frac{\pi}{2}$. Panel (b): the same evolution on the Bloch 
sphere.  The decoherence parameter is set to $\mu = 0.1$ and the 
natural frequency to $\omega = 1$.}
\label{fig:sferamil2liv}
\end{figure}

We solve Eq. (\ref{mil}) for the free evolution of a qubit, $ H= \omega \sigma_{3}$, using
the Bloch representation $\rho= \frac{1}{2} ( \mathbb{1} +\bar{r}  \cdot \bar{\sigma} )$, 
where $\bar{r} = \{r_1,r_2,r_3\}$. We have 
\begin{equation}
\begin{array} {lll}
 \dot{r_1}  = - 2 \omega r_2 - 4 \mu \omega^2 r_1   \\
 \dot{r_2} =  2 \omega r_1 - 4 \mu \omega^2 r_2  \\
 \dot{r_3} = 0
\end{array} \,.
\label{sistemaeqdiffmil}
\end{equation} 
For a generic pure initial state,  {\em i.e.}, $ r_{10} = \cos{\phi} \sin{\theta},  r_{20} = \sin{\phi} \sin{\theta},  r_{30} = \cos{\theta},$
we obtain
\begin{equation}
\begin{array} {lll}
 r_1(t)  = e^{-4\mu \omega^2 t} \cos(2\omega t + \phi) \sin{\theta} \\
 r_2(t) =  e^{-4\mu \omega^2 t} \sin(2\omega t + \phi) \sin{\theta} \\ 
 r_3(t) = \cos{\theta}
\end{array} \,.
\label{radius}
\end{equation} 
Eq. (\ref{radius}) says that $r_1$ and $r_2$ remain on the same horizontal 
plane of the Bloch Sphere since $r_3$ is constant. On this plane, $r_1$ 
and $r_2$ oscillate and vanish exponentially. The constancy of $r_3$ 
indicates that the ratio between $\ket{0}$ and $\ket{1}$ in the 
statistical mixture remains constant, while coherence vanishes,  
{\em i.e.}, the superposition turns into a statistical mixture. 
The larger $\omega^2$ is, the quicker is decoherence. In Fig. (\ref{fig:sferamil2liv}), we show the evolution of the Bloch vector 
starting from $\theta= \frac{\pi}{4}$ and $\phi= \frac{\pi}{2}$ 
for $\omega = 1$. Notice that the decoherence parameter has been 
set to $\mu = 0.1$, a value significantly larger than the expected 
one, to enhance graphic clarity. In the standard basis, Eq. (\ref{mil}) 
rewrites as 
\begin{equation}
 \dot{\rho}= 
\begin{pmatrix} 
0 & -2i\omega -4 \omega^2 \mu  \\
2i\omega - 4 \omega^2 \mu  & 0\\
\end{pmatrix} \,.
\label{matreq}
\end{equation}
and the solution for a generic initial condition is given by 
\begin{equation}
\rho(t)= 
\begin{pmatrix} 
\cos{\frac{\theta}{2}}^2 & \frac12 \eta_m \sin{\theta}\, e^{-i\phi}  \\
\frac12 \eta_m^*  \sin{\theta}\,
e^{i\phi}  \ & \sin{\frac{\theta}{2}}^2\\
\end{pmatrix} 
\label{matricetemporalemilburn}
\end{equation} 
where $\eta_m=e^{-2i \omega t -4 \mu \omega^2 t}$, and we can 
explicitly observe that the non-diagonal matrix elements, 
representing phase coherence, vanish for increasing $\omega t$, 
meaning that the system is experiencing dephasing.
In order to assess the estimability of the decoherence parameter 
we now proceed by calculating the quantum Fisher information. 
For a two-level system with states labeled by 
the parameter $\lambda$, the quantum Fisher information may be 
evaluated in terms of the Bloch vector as follows \cite{11} 
\begin{equation}\label{nlab}
       Q(\lambda) = |\partial_{\lambda} \bar{r}|^2+ \frac{1}{1-|\bar{r}|^2}(\bar{r} \cdot \partial_{\lambda}\bar{r})^2\,.
\end{equation} 
Using the expression in Eq. (\ref{radius}), we find 
\begin{equation}
    Q(\mu)= \frac{16\, \omega^4 t^2 \sin^2{\theta}}{e^{8\mu \omega^2 t}-1}.
\end{equation}
If we then consider the quantum signal-to-noise ratio $R(\mu)$, 
it is apparent that it
depends solely on the variables $x \equiv \mu \omega^2 t$ and $\theta$. 
Notably, this function is maximized when $\theta = \frac{\pi}{2}$. 
This observation is consistent with the fact that maximum coherence 
is achieved at this angle, leading to the most significant deviation 
of the MID temporal evolution from the unitary one. Similarly, 
it is not surprising that $Q(\mu) = 0$ when $\theta = 0 \vee \pi$ 
since $\mu$ does not induce any variation in the evolution of the 
eigenstates of $H$. We then need to analyze the single-variable 
function $R = \frac{16 x^2}{e^{8x}-1}$, which is maximized at 
$x \simeq 0.199$, corresponding to $R \simeq  0.162$.  
\par
\par
Overall, we have that for any value of $\mu$,  it is possible 
to select an optimal value of $\omega^2 t$ so that the 
\textit{signal-to-noise ratio} is maximized and maintains the same 
value. This possibility is a promising outcome, indicating 
that even very small values of the decoherence parameter may be 
reliably estimated.

The QFI establishes the upper bound on precision over all 
possible POVMs. However, it is not guaranteed that the POVM 
saturating the bound is a practically feasible one. In order to 
investigate  whether it is possible to achieve the maximum precision 
using a spin measurement, we evaluate the Fisher Information 
(\ref{Fisher}) of the 
generic spin measurement $\left\{ \ket{\theta\phi}\bra{\theta\phi}, \mathbb{I}- \ket{\theta\phi}\bra{\theta\phi}\right\}$ where  $\ket{\theta\phi} = \cos{\frac{\theta}{2}} \ket{0}+ e^{i \phi} \sin{\frac{\theta}{2}} \ket{1}$, and $\ket{0}$ and  $\ket{1}$ are eigenstates of $\sigma_3$.
We obtain 
\begin{equation}
  F = -\frac1S \left(64 t^2 \omega^4 \cos^2 2\omega t \sin^2{\theta}\right)
\end{equation}
where $S=  
1 - 2 e^{8 t \mu \omega^2} (3 + \cos{2 \theta}) 
+ 8 e^{4 t \mu \omega^2} \cos^2{\theta} \cos2\omega t 
- 2 \cos{2 \theta} \cos^2 2\omega t  + \cos4\omega t $.
By maximizing over $\theta$, we find the optimal spin measurement for 
$\theta= \frac{\pi}{2}$. We then consider the ratio between the optimal FI 
and the QFI
\begin{equation}
    \frac{F}{Q}= -\frac{2 \left( -1 + e^{8 t \mu \omega^2} \right) \cos^22\omega t }{
1 - 2 e^{8 t \mu \omega^2} + \cos4 \omega t}
\end{equation}
and observe that it only depends on the two adimensional 
variables $y= \mu \omega $ and $x= \omega t$. 
For $x=\pi/2$ the ratio is equal to $1$ for any $y$. The FI is then maximized 
for $y \simeq 0.127$, which corresponds to the optimal setting to estimate 
$\mu$. We conclude that, at least in principle, the MID model is amenable to 
falsification through a spin measurement.
\subsection{Harmonic oscillator}
Here we apply the MID model to a free harmonic oscillator, {\em i.e.},  
$H= \omega  a^\dagger a$, and analyze whether and under which conditions 
this system is suitable for estimating the value of the dephasing parameter.
We start by writing Eq. (\ref{mil})  in the Fock basis 
\begin{equation}
\dot{\rho}_{nm}  = -i\omega(n-m)  \rho_{nm} - 
 \mu \omega^2 (n-m)^2 \rho_{nm}
 \label{masteroscill}
\end{equation}
(dot denotes time derivative) which is solved  by
\begin{equation}
   \rho_{nm}(t) = C_n(0) C_m^*(0) e^{-i\omega(n- m)t-\mu{ \omega^2 (n-m)^2t}}
\end{equation}
where $C_n(0)$ is the amplitude of the initial state along 
the $m$-th eigenstate of the Hamiltonian and $C_m^*(0)$ denotes the complex conjugate.
\mga{In particular, we consider the oscillator initially prepared in a coherent state,  {\em i.e.}, 
\begin{align}
      C_{nm}(0)& =  e^{-|\alpha|^2} \frac{\alpha^n \alpha^{*m}}{\sqrt{n!m!}}\,,
\end{align}
and use Eq. (\ref{numerico}) to evaluate the QFI. 
More specifically, we consider the evolved density matrix, with elements
\begin{equation}
\rho_{mn}=  \frac{\alpha^n \alpha^{*m}}{\sqrt{n!m!}} 
e^{-|\alpha|^2} e^{-i\omega(n-m)t} e^{-\mu \omega^2 (n-m)^2 t}
\label{matrixelement}
\end{equation}
and diagonalize it by truncating the Hilbert space at a suitable dimension to ensure normalization. }
\begin{figure}[ht!]
    \centering
    \includegraphics[width=0.52\columnwidth]{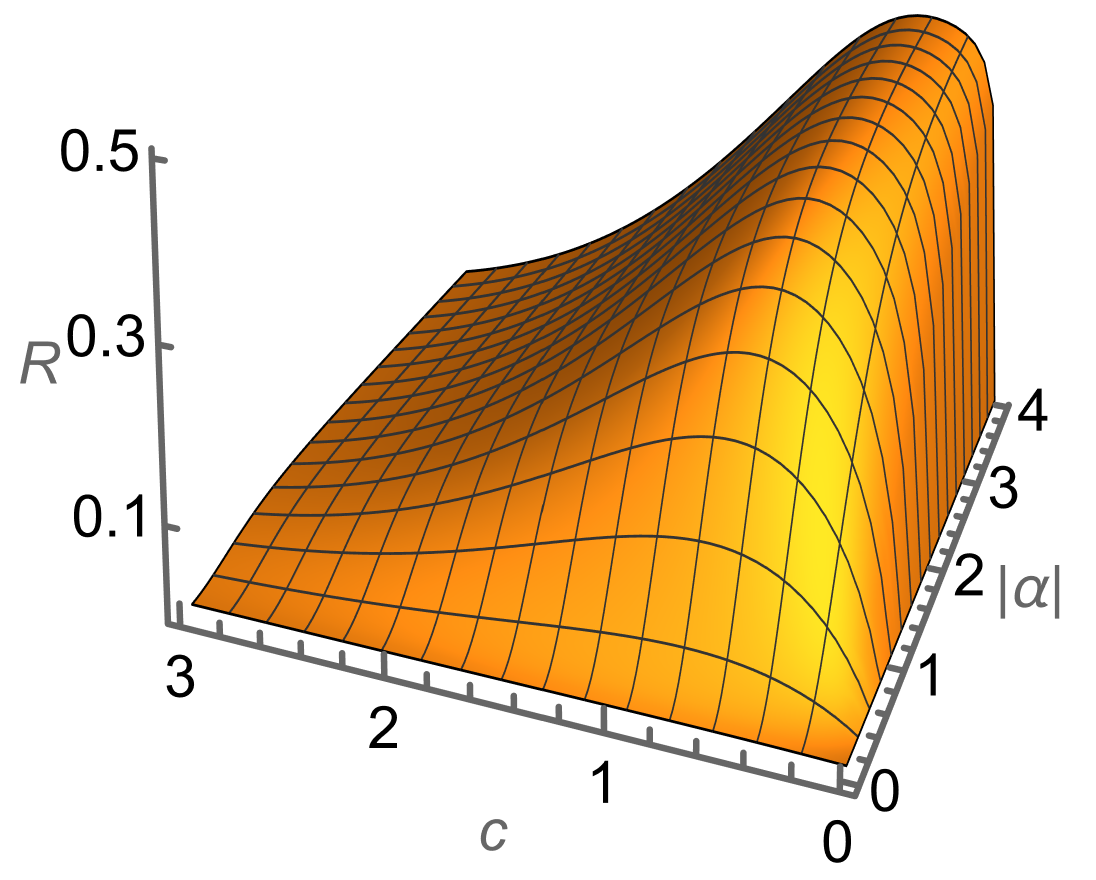} 
     \includegraphics[width=0.45\columnwidth]{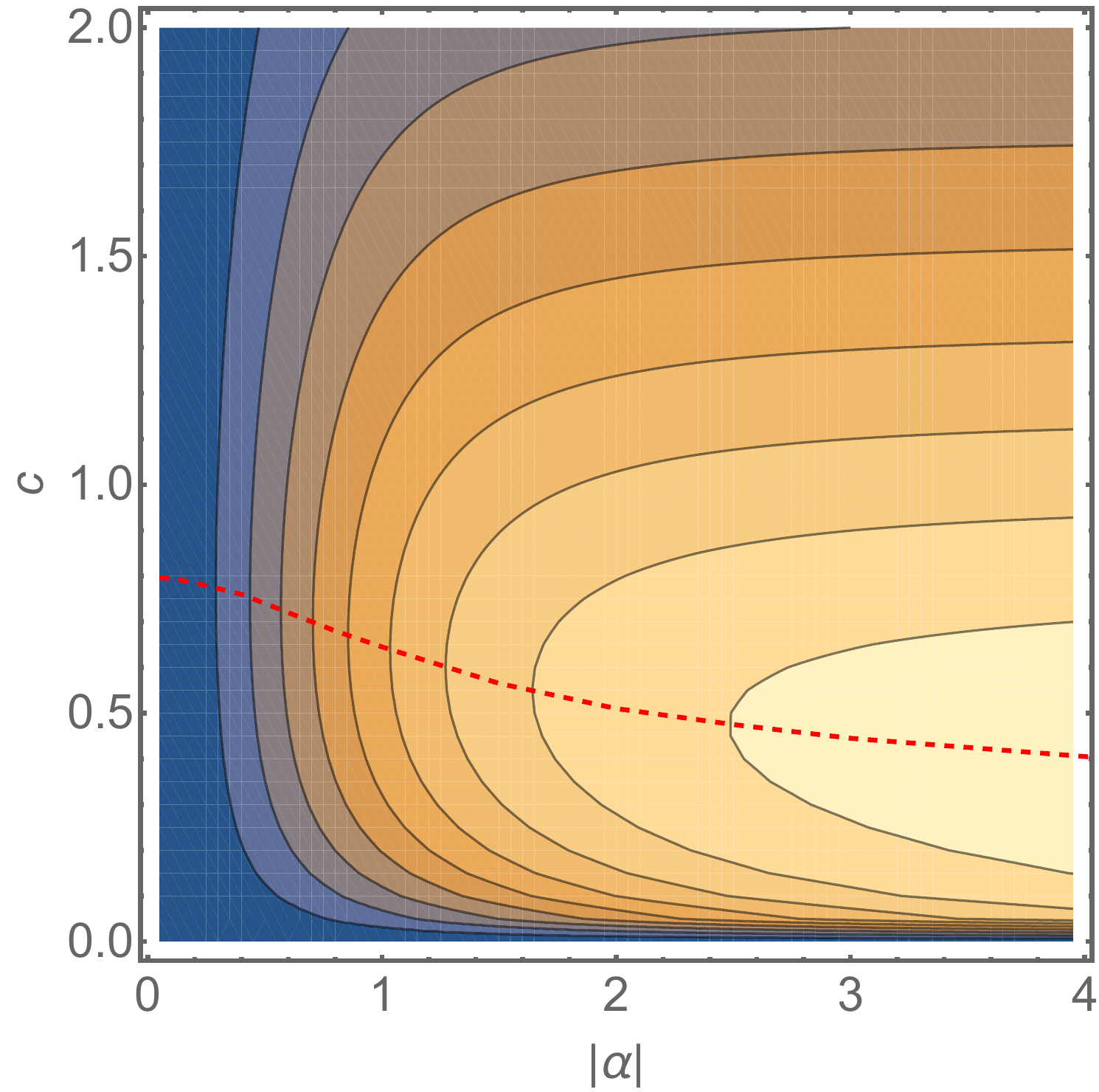} 
      \includegraphics[width=0.48\columnwidth]{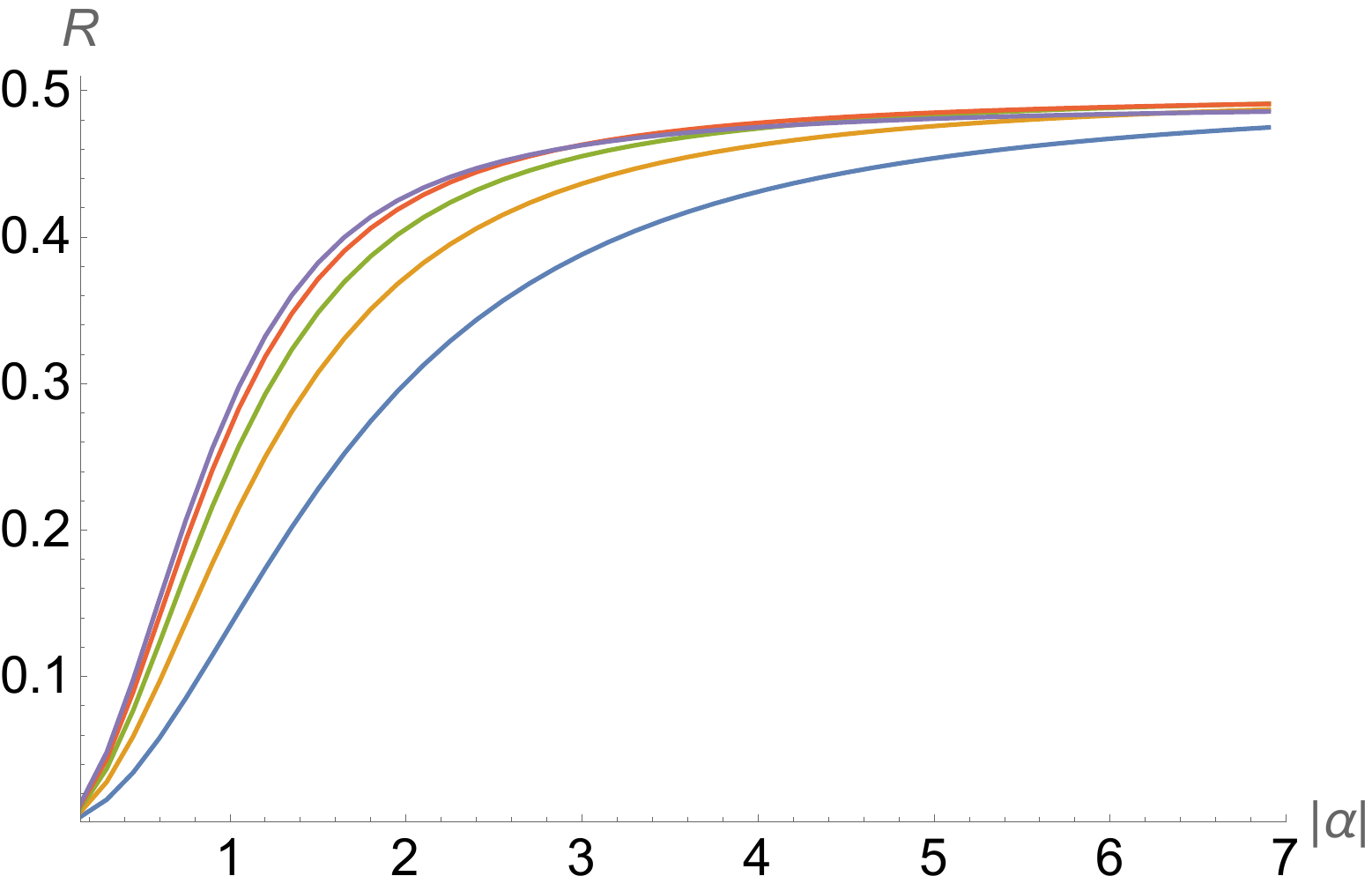} $\,\,$ 
       \includegraphics[width=0.47\columnwidth]{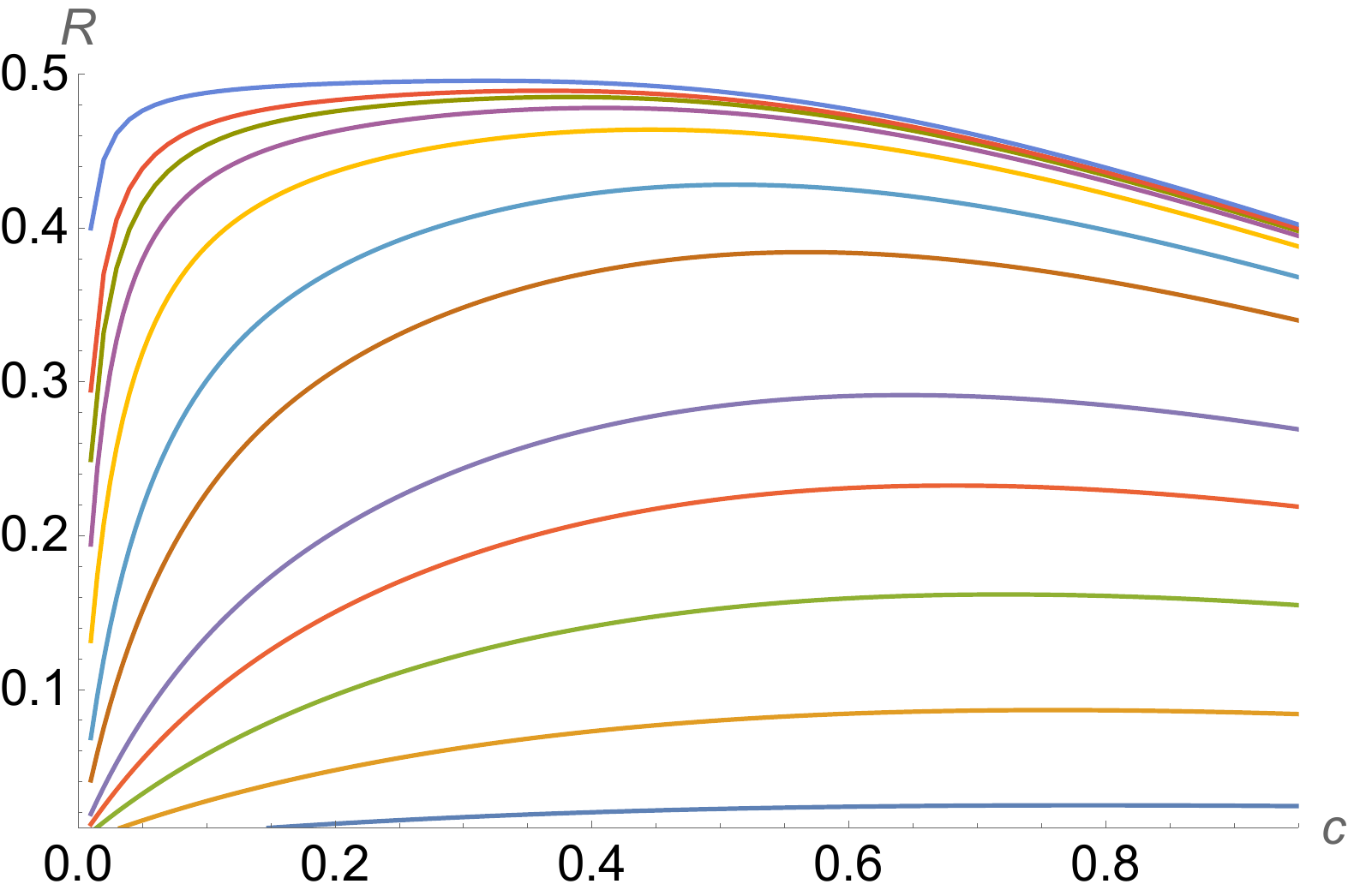} 
    \caption{\mga{The quantum signal-to-noise ratio $R$ for the estimation of the decoherence parameter of a harmonic oscillator system, $H= \omega a^{\dagger} a$, evolved according to the MID model. 
    The QSNR depends on the coherence amplitude $\alpha$ and the adimensional variable $c= \mu \omega^2 t$ and it is an increasing function of $|\alpha|$. In the left upper panel we show $R$ as function of $\alpha|$ and $c$, whereas in the right upper panel we show the same function as a contour plot.
    At fixed value of the coherence amplitude the maximum is achieved for $c_m=g(|\alpha|)$, where $g$ is a (slowly) decreasing function of $|\alpha|$, denoted by the red dashed line. For $0\leq |\alpha| \leq 10$ a best fit for $g$ gives $g(x) = 0.32 + 0.51 e^{-0.45 x}$. The asymptotic value of $R$, obtained for large $|\alpha|$ and $c=c_m$, is $R=1/2$. In the lower panels, we show $R$ as a function of $|\alpha|$ for different fixed values of $c$ (from bottom to top the curves are for $c=0.1,0.2,0.3,0.4,0.5$, respectively), and as a function of $c$ for different fixed values of $|\alpha|$ (right, from top to bottom the curves are for $|\alpha|=10,6,5,4,3,2,1,0.5,0.4,0.3,0.2,0.1$, respectively).}}
        \label{fig:qfimilosc}
\end{figure}

\mga{In order to maximize the QFI, we look for its 
scaling properties. An extensive 
numerical analysis, corroborated by analytic considerations, reveals 
that the QFI is a function of two variables only, the coherent 
amplitude $\alpha$ and the adimensional variable $c= \mu \omega^2 t$. In the upper panels of Fig. 
\ref{fig:qfimilosc}, we show the behavior of the corresponding QSNR as a function of these two variables. R is an increasing function of the coherent amplitude, whereas it shows a maximum as a function of $c$, located at $c_m = g(|\alpha|)$, where $g$ is a (slowly) decreasing function of the coherent amplitude. 
The function $g(|\alpha|)$ is the red line in the right upper panel of Fig.  
\ref{fig:qfimilosc}.
For $0 < |\alpha| < 10$, $g$ may be approximated by $g(x) = 0.32 + 0.51 e^{-0.45 x}$. The value of the maximum increases with $|\alpha|$ and approaches $R=1/2$ for large $|\alpha|$.  For $c$  values near $c_m$, $R$ reaches its asymptotic value even at relatively small amplitudes.}
\section{Metrology of the GND parameter} \label{s:diss}
Let us now address the non-linear, deterministic, dissipative model suggested by Gisin (GND model) \cite{9} and discuss whether the value of the intrinsic dissipation parameter $\gamma$ may be inferred by monitoring a two-level system or a harmonic oscillator.
\subsection{Two-level system}
\begin{figure}[ht!]
\centering
\includegraphics[width=0.44\columnwidth]{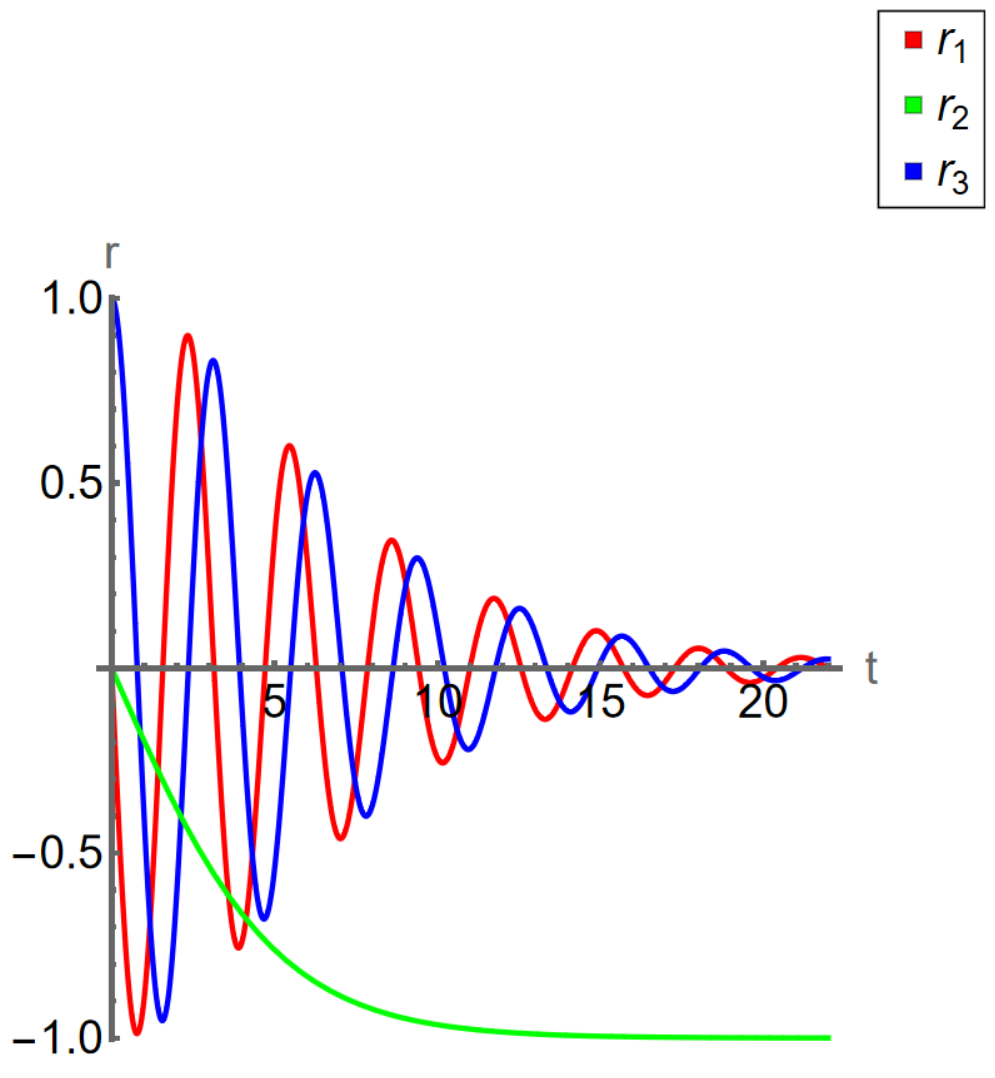}
\includegraphics[width=0.44\columnwidth]{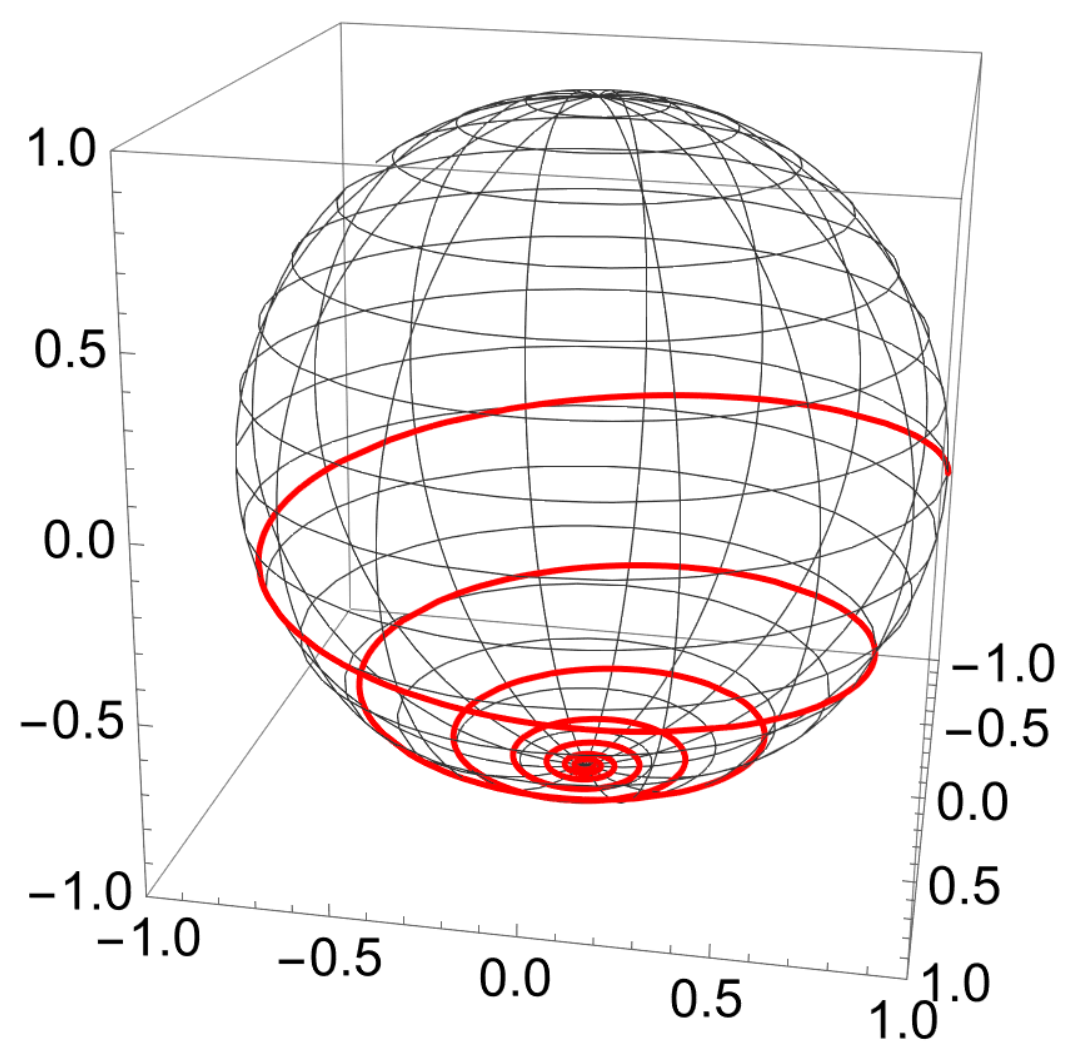}
\caption{Panel (a): Temporal evolution of the components of the Bloch vector 
 of a two-level system with $H= \omega \sigma_3 $, according to GND model, with initial condition $\theta= \frac{\pi}{2}$ and \md{$\phi=\frac{\pi}{2}$}. Panel (b): the same evolution 
 on the Bloch sphere. The decoherence parameter is set to $\gamma = 0.1$ and the natural frequency to $\omega = 1$.}
\label{fig:gisin2livsfera}
\end{figure}

We start from the master equation (\ref{mastergisin}) and write it in the Bloch sphere representation
\begin{align}
 \dot{r_1}  & = - 2\omega r_2 + 2 \gamma \omega r_1 r_3  \\
 \dot{r_2} & = 2 \omega r_1 + 2 \gamma \omega r_2 r_3  \\
 \dot{r_3} & = - 2 \gamma \omega( r_1^2 +r_2^2)\,.
\label{sistemaeqdiff}
\end{align}
Since in the GND model purity is preserved, if we start from a pure state we have $ \left | r \right | = 1 $ at any time. Consequently, the third equation can be 
rewritten as $\dot{r}_3 = - 2\gamma \omega (1 -r_3^2 ) $, with solution:
\begin{equation}
    r_3(t)= -1 + \frac{2}{D}(1+ \cos{\theta})
\label{r_3}
\end{equation}
where $ D \equiv 1- e^{4t\gamma \omega} (-1 + \cos{\theta}) + \cos{\theta}$. Inserting, 
Eq. (\ref{r_3}) in Eq. (\ref{sistemaeqdiff}) and assuming a generic initial condition, 
we arrive at 
\begin{align}
r_1(t) & = \frac{2}{D}\,  e^{2\gamma \omega t}\cos(\phi + 2 \omega t) \sin{\theta}  \\
r_2(t) & = \frac{2}{D}\,  e^{2\gamma \omega t} \sin(\phi + 2 \omega t) \sin{\theta}
\label{r_1,2}
\end{align}
This shows that $r_1$ and $r_2$ oscillate and vanish exponentially,  {\em i.e.}, coherence
decreases with time. Meanwhile, $r_3$ approaches -1,  {\em i.e.}, energy is dissipated 
and the state evolves toward the lowest energy level. Finally, the modulus of the Bloch vector $|\bar{r}|=1$ is constant and remains on the Bloch Sphere surface,  {\em i.e.} purity is preserved.  In Figure (\ref{fig:gisin2livsfera}), we illustrate the evolution of the 
Bloch vector for initial conditions $\theta= \frac{\pi}{2}$ and $\phi= \frac{\pi}{2}$. The decoherence parameter is 
set to $\gamma = 0.1$, a value significantly larger than the expected one, chosen  to enhance graphic clarity. We have also set $\omega = 1$.

The evolved density matrix can be written as \begin{equation}
\rho= 
\begin{pmatrix} 
\left[1+ e^{4 \gamma \omega t }\tan^2{\frac{\theta}{2}}\right]^{-1} & \eta_g\, D^{-1}\,\sin{\theta} \, e^{-i\phi}\\ & \\
\eta_g^*\,D^{-1}\,\sin{\theta}\, e^{i\phi}& 1-\left[1+ e^{4 \gamma \omega t }\tan^2{\frac{\theta}{2}}\right]^{-1} 
\end{pmatrix} \,,
\label{matricetemporale}
\end{equation} 
where $\eta_g = e^{-2(i -  \gamma) \omega t}$, and we can clearly see the lowest energy level being filled with time. 

Since the system is pure,  Eq. (\ref{nlab}) reduces to $Q(\lambda) = \left | \partial_{\lambda} \bar{r} \right |^2$ \cite{11}
leading to
\begin{equation}
    Q(\gamma)= 16 \omega^2 t^2 \, \frac{e^{4\gamma \omega t} \sin^2{\theta}}{(1- e^{4t\gamma \omega} (-1 + \cos{\theta}) + \cos{\theta})^2}\,.
\label{r_gbl}
\end{equation}
The quantum signal-to-noise ratio may be written in terms of the state parameter $\theta$ and of the adimensional variable $x = \gamma \omega t$. 
\begin{equation}
      R(x,\theta)= 16 x^2\frac{e^{4x} \sin^2{\theta}}{(1- e^{4x} (-1 + \cos{\theta}) + \cos{\theta})^2}\,.
\end{equation}
\mga{The maximum is found for 
$
   \cos{\theta_m} = (e^{4 x t }-1)/( e^{4 x t}+1) $}
which corresponds to the value $R= 4 (x)^2$. \md{Given that $R$ is a monotonically increasing function of $x$ , it is advisable to select $x$ as large as experimentally feasible. Under this condition, the maximum of  $R$ is obtained when $\theta$ approaches zero.}
\md{However, special attention must be given to the discontinuity in the QFI at $ \theta = 0$ , where $Q(\gamma) = 0$. This discontinuity underscores a physical aspect: at $ \theta = 0$,  the eigenstates remain stationary under this type of evolution, reflecting the lack of dynamics necessary to encode measurable information about the parameter. }

\mga{Let us now evaluate the FI of a generic spin measurement. Following the steps already followed in the previous Section we arrive at 
\begin{align}
F= -\frac{4 \tau}{B}  \Big[ \sin2 \theta - 
A \cosh 2 \tau   \Big]^2
\end{align}
where $\tau=\omega t$
$ A= \cosh 2 x \sin 2 \theta 
 - 2 \sin\theta \sinh 2 x$ and 
 $B  =  2 [\cos2 \tau - \cosh 2x]  (3 + \cos 2 \theta
\cosh 2 x + \, 4 \cos 2  \tau \sin^2 \theta - 8 \cos\theta \sinh 2 x  [ \cosh 2 x 
-  \cos\theta \sinh 2 x]^2$.}
Upon choosing $\theta=\theta_m$ and $\omega t= k \frac{\pi}{2}$, we have $F=Q$ independently of $\gamma$. However, in order to make $R$ independent of $\gamma$, large values of time $k= \frac{1}{\gamma}$ should be chosen. If this constraint may be fulfilled experimentally, then it would be possible to falsify the GND model through spin measurements. 
\subsection{Harmonic oscillator}
According to Eq. (\ref{formal}), the evolution of a generic state of a harmonic 
oscillator according to the GND model can be written as 
\begin{equation}
    \ket{\psi_t} = \sum_n \frac{e^{-(i+ \gamma ) \omega n t} C_n(0) \ket{n}}{\sqrt{ \sum_n |C_n(0)|^2e^{- 2\gamma \omega n t} } }.
\label{psitgisin}
\end{equation}
Starting from a coherent state $|\alpha\rangle$, we obtain that the evolved state is still coherent with a decreased amplitude $|\alpha\rangle e^{-\gamma\omega t}$.  \md{As $ t \rightarrow\infty$, the system is brought towards its lowest energy configuration.}

\md{Since the state is pure, the QFI  can be calculated using Eq. (\ref{QFIpuri}). }\md{For this model it corresponds to $
    Q(\gamma)= 4t^2[ \mel{\psi_t}{H^2}{\psi_t} - \mel{\psi_t}{H}{\psi_t}^2]$, which is consistent with results presented in previous works  \cite{BrodyGisin} }
\begin{equation}
    Q(\gamma) = 4\omega^2 t^2 |\alpha|^2 e^{-2\gamma \omega t}\,.
\end{equation}
\mga{The quantum signal-to-noise ratio $R(\gamma)$ is a function of the sole variable $x= \gamma \omega t$ and is maximized for $x=1$, corresponding to 
$R=4|\alpha|^2 $. This result not only indicates that the dissipation parameter can be estimated regardless of its specific value, but it also demonstrates that the estimation precision can be further enhanced by increasing the energy of the probe state}.  
\section{Discussion and conclusions}
\label{s:out}
In Table \ref{tab:my-table} we summarize the optimal conditions required to maximize the estimability of the intrinsic decoherence parameters of MID and GND models, using a two-level system or a harmonic oscillator.
\begin{table}[h!]
\centering
{%
\mga{\begin{tabular}{|c|c|c|}
\hline
 & \textbf{MID model} & \textbf{GND model} \\
 \hline
\multicolumn{1}{|c|}{
Two-level} &
  \begin{tabular}[c]{@{}c@{}}
    $ t \simeq 0.199/ (\mu \omega^2)$ \\
    $\theta= \frac{\pi}{2}$ \\
    $R\simeq 0.162$
  \end{tabular} &
  \begin{tabular}[c]{@{}c@{}}
    $t \gg 1/\omega$ \\
    $\theta= \arccos \left( \frac{e^{4 \gamma \omega t} - 1}{e^{4\gamma \omega t} + 1} \right)$ \\
    $R= 4 (\gamma \omega t)^2$
  \end{tabular} \\ \hline
\multicolumn{1}{|c|}{Oscillator} &
  \begin{tabular}[c]{@{}c@{}}
    $ t= g(\alpha)/(\mu\omega^2)$ \\
     $\alpha \gg 1$ \\
    $R =1/2$
  \end{tabular} &
  \begin{tabular}[c]{@{}c@{}}
    $t= 1/(\gamma\omega)$ \\
    $\alpha \gg 1$ \\
    $R= 4 |\alpha|^2$
  \end{tabular} \\ \hline
\end{tabular}}
}
\caption{Summary of the conditions to maximize the estimability of the intrinsic decoherence parameters of MID and GND models.}
\label{tab:my-table}
\end{table}

\mga{The differing behaviors of the QSNR are a natural consequence of the fundamental differences between the two models. For a two-level system 
in the MID model, the optimal probe state is fixed, 
corresponding to a maximumally coherent state, and independent of evolution 
time, meaning that precision is not significantly enhanced by extending the evolution. Conversely, in the GND model, both the optimal state and QSNR depend on the available time, with longer evolution enabling better exploitation of the GND model encoding and improved precision. The ultimate optimal state for the GND model is close to the north pole of the Bloch sphere, but its potential precision may be achieved only with sufficient evolution time.
For the harmonic oscillator and both models, precision improves as 
the coherent amplitude $\alpha$ increases. Moreover, in the GND model, this leads to a further enhancement of the QSNR, consistent with the idea that 
having large energy enables more effective detection of dissipation.}

We should notice that for both systems and both models, since the expected values 
of $\mu$ or $\gamma$ are small, the optimal conditions require a large value of the interaction time,  {\em i.e.},  $t\gg 1/\omega^2$ for MID model and $t\gg 1/\omega$ for 
the GND one. However, one should also take into account the intrinsic decoherence 
phenomena are expected to occur in a shorter time scale compared to extrinsic 
ones \cite{decoherencehistories} and by performing 
experiments with an excessively large $t$ we may/should attribute dephasing and 
dissipation of our system to the interaction with the external environment, 
rather than to the temporal evolution, making the experiment inconclusive. 
A possible solution is to employ systems characterized by a large value of 
$\omega$. 

Another concern may arise about the optimal values of the evolution time, which in 
principle 
does depend on the value of the decoherence parameters, which is unknown and is just 
what we are trying to estimate. We can address this issue by employing an iterative strategy where we initially guess a reasonable value of the parameter, perform a measurement in 
the corresponding optimal conditions, and use the estimated value of the parameter 
to determine renewed and improved conditions. As these conditions should yield a smaller variance, we anticipate obtaining a more accurate value of the parameter for the subsequent step. This iterative process continues, and if we begin with a reasonable value 
of the decoherence parameter, we should eventually converge to a satisfactory 
level of precision.

In the cases we examined, the quantum signal-to-noise ratio can remain constant (as in the MID model) or even increase over time or with probe energy (as in the GND model). 
This is a promising outcome since it suggests that with a sufficiently large number of measurements, it is possible to estimate the IDM 
parameter with a suitable precision, in turn making the intrinsic decoherence model falsifiable and worth studying to assess their validity. In conclusion, our primary aim has been to assess the testability of two paradigmatic models of intrinsic decoherence. Our results demonstrate that there are optimal conditions under which the MID and GND models can be falsified,  {\em i.e.}, it is possible to conduct experiments to validate or refute those models. Simply put, IDMs {\em may be wrong} after all \cite{noteven}.

\acknowledgements{The authors sincerely thank Jakub Rembieli\'nski for engaging discussions and insightful suggestions. They also extend their gratitude to Dorje C. Brody and the anonymous Referees for their valuable feedback and recommendations.}

\bibliographystyle{eplbib}
\bibliography{idmbib}
\end{document}